# Secure Spectrum and Resource Sharing for 5G Networks using a Blockchain-based Decentralized Trusted Computing Platform.

[1]Hisham A. Kholidy, *Senior Member, IEEE,* [2]Mohammad A. Rahman, *Senior Member, IEEE,* [3]Andrew Karam, [4]Zahid Akhtar, *Senior Member, IEEE*

[1]Department of Networks and Computer Security, College of Engineering, SUNY Polytechnic Institute, Utica, NY, USA.
kholidh@sunypoly.edu

[2]Florida International University, Department of Electrical and Computer Engineering, USA
rahmanma@sunypoly.edu

[3]RIGB, The Air Force Research Lab(AFRL) Rome, NY,USA
andrew.karam@us.af.mil

[4]Department of Networks and Computer Security, College of Engineering, SUNY Polytechnic Institute, Utica, NY, USA.
akhtarz@sunypoly.edu

**Abstract**

The 5G network would fuel next-gen, bandwidth-heavy technologies such as automation, IoT, and AI on the factory floor. It will improve efficiency by powering AR overlays in workflows, as well as ensure safer practices and reduce the number of defects through predictive analytics and real-time detection of damage. The Dynamic Spectrum Sharing (DSS) in 5G networks will permit 5G NR and 4G LTE to coexist and will provide cost-effective and efficient solutions that enable a smooth transition from 4G to 5G. However, this increases the attack surface in the 5G networks. To the best of our knowledge, none of the current works introduces a real-time secure spectrum-sharing mechanism for 5G networks to defend spectrum resources and applications. This paper aims to propose a Blockchain-based Decentralized Trusted Computing Platform (BTCP) to self-protect large-scale 5G spectrum resources against cyberattacks in a timely, dynamic, and accurate way. Furthermore, the platform provides a decentralized, trusted, and non-repudiating platform to enable secure spectrum sharing and data exchange between the 5G spectrum resources

1. **Introduction**

A 5G network would dramatically improve military communication and situational awareness. One of the 5G capabilities called Dynamic Spectrum Sharing (DSS), enables a radio channel to support both 4G LTE and 5G and allows the 5G signal to share the underutilized spectrum. Considering the traffic characteristics of current cellular network communications and the spectrum shortage dilemma, the DSS is a cost-efficient solution. [1] The DSS will permit 5G NR and 4G LTE to coexist, allowing network operators a smooth transition from LTE to 5G NR. With DSS, 5G can share the frequency bands used by 4G, allowing operators to swiftly expand 5G coverage while eliminating the need for new spectrum allocation for 5G. [1] Despite this, the implementation of spectrum sharing in 5G heterogeneous networks is fraught with challenges, such as security threats, privacy leakage, and lack of incentive mechanisms. [2, 3] Threats against spectrum sharing exploit either spectrum sensing or geo-location databases. [4] According to [3, 4], we can classify these threads into two broad categories: threats to sensing-driven spectrum sharing and threats to database-driven spectrum. The first category includes threats that affect (1) the physical layer. E.g., the Primary User Emulation (PUE) attack [5], (2) the MAC layer. E.g., A rogue transmitter may corrupt the cognitive control channel [6] leading to a DoS attack, beacon falsification (BF) attack [7], and small-back-off-window (SBW) attack [8]. (3) cross-layer where attacks can be conducted concurrently to exploit vulnerabilities in two or more layers. E,g, Lion attack [9] that targets the physical and transport layers of a Cognitive Radio Network (CRN). The second category includes two subclasses: (1) database inference attacks. This subcategory includes (a) threats to the privacy of primary users which is an especially critical concern related to the sharing of the federal government (including military) spectrum with non-government systems. [10] (b) threats to the privacy of Secondary Users (SUs). E.g., Spectrum Utilization-based Location Inferring attack [11]. (2) Threats to the database access protocol such as security concerns related to using a geolocation database, impersonating a certified device, jamming a query or a database response...etc.

Blockchain is a technology of collectively maintaining a reliable database through decentralization and de-trust. There are several applications for Blockchain technology in the security domain [12]. [13] proposed a new type of distributed cloud architecture based on blockchain that provides secure and on-demand access to the most competitive computing infrastructure in the Internet of Things (IoT). [14] proposed a multi-layer security model of IoT based on blockchain at all levels of the network and provides a wide-area networking solution of the Internet of Things. [15] proposed a blockchain-based smart home framework. Each smart home is equipped with an always-on, high-resource device called a "miner" that handles all communications inside and outside the home. [16] combined blockchain with a big data system and proposed a method to overcome the shortcomings of existing identity verification through blockchain technology. [17] proposed a blockchain-based traffic network architecture (Block-VN) in smart cities.

To the best of our knowledge, none of the current works introduces a real-time secure spectrum-sharing mechanism for 5G networks to defend spectrum resources and applications. This paper aims to propose a Blockchain-based Decentralized Trusted Computing Platform (BTCP) to self-protect large-scale 5G spectrum resources against cyberattacks in a timely, dynamic, and accurate way. Furthermore, the platform provides a decentralized, trusted, and non-repudiating platform to enable secure spectrum sharing and data exchange between the 5G spectrum resources. This platform can effectively bridge the gap between spectrum supply and demand in 5G networks and reduce the cost. The BTCP ensures that every spectrum transaction can be safely and reliably saved in the block and does not need to pay for a third-party platform. In addition, it can ensure that free spectrum can be shared in real time and efficiently, thus saving the operator's cost and improving the overall spectrum utilization at the same time.

1. **Literature Review**

[18] Wajdi et al studied the spectrum sharing security from the cognitive radio MAC layer perspective. Authors classified the threats against cognitive radio into two different categories.

The first category includes threats specific to CRN users namely (1) threats against spectrum sensing that bring the network performance down by reporting the false results of the Primary User (PU) detection. (2) threats against spectrum management such as the forgery or tampering attacks that are designed to transmit incorrect spectrum sensing information to the data collection center to deceive the secondary user. (3) threats against Spectrum Mobility that force SUs to vacate the channel by pretending to be the PU. (4) threats against Spectrum sharing that includes the abovementioned two broad threats categories in Section 1 (i.e., threats to sensing-driven spectrum sharing and threats to database-driven spectrum). The second category relates to common security threats in both conventional wireless and CR networks. This category includes fake attacks, information tampering, service repudiation, replay attack, denial of service and information, interference, greedy behavior attack, malicious and selfish behavior attacks, and black and grey hole attacks.

In [19], Fan et al. proposed a generalized temporal-spatial spectrum sharing scheme for ultra-dense networks (UDNs) by using a game-theoretical approach. In [20], Kumar et al. presented a cooperative spectrum sharing framework and designed a novel dynamic resource allocation strategy based on fractional frequency reuse. A new network architecture was proposed for the coexistence of WiFi and cellular systems [21] based on the almost blank subframe (ABS) protocol.

Yu et. al. [22] proposed a spectrum sharing scheme in which a designated spectrum broker dynamically allocates spectrum to a set of secondary networks with different access priorities. Game-theoretic approaches for dynamic spectrum leasing in cognitive radio networks have been studied in [23, 24]. In [24], the PU encourages or discourages the spectrum allocation to SUs depending on their maximum power constraints and Quality-of-Service (QoS).

The current state of the art relies on a centralized agency to verify every spectrum sharing transaction. This makes these mechanisms nonscalable because of the exponential growth of the spectrum allocation problem and their existing vulnerabilities to DoS attacks and a single point of failure problem.

[25] Moreover, they mainly consider resource management and neglect the critical features for spectrum sharing such as security and privacy issues. Other approaches use the traditional incentive methods that are based on the assumption that the Human-to-Human (H2H) [26] side information is already known to the Base Station (BS), which might be impractical in a real-world application. Therefore, the development of an incentive mechanism with non-symmetrical information is another real challenge.[26]. In addition to that, the current incentive mechanisms have not considered the mutual preferences of H2H users and Machine-Machine (M2M) users. The H2H users may not have the motivation to share their spectrum resources without a motivating compensation [26] and did not introduce a real-time solution to address the authentication in decentralized CRNs.

## 2. The Proposed Blockchain-based Decentralized Trusted Computing Platform (BTCP).

While 5G Edge enables a dynamic edge environment, the participants are not fixed, and they are heterogeneous in terms of their spectrum resources and 5G Edge nodes. Moreover, some edge nodes may advertise to offer a certain spectrum, edge computing, and storage capacities but may provide different performance behavior during the participation. Hence, there is a need for accountability/reputation, along with incentives for quality participation.

The BTCP addresses the security threats that are inherited from the spectrum sharing aforementioned above. It revamps our existing autonomous security framework [27-53] with trust management, and modeling and analysis techniques and leverages our 5G security testbed [27, 29] to provide a lightweight Blockchain-based decentralized trusted framework that is essential to (1) provide reliable privacy and security guarantees for spectrum sharing (2) Maintain anonymity from other participants using a trust management mechanism.

## 3. Methods and Procedures

The BTCP is a decentralized trusted computing framework that manages the participating edge devices using the non-interactive zero-knowledge proof (ZKP) technology. The zkp is a perfect candidate to be leveraged to ensure authentication, protecting participating 5G Edge node' (corresponding users' or UE') privacy. The Blockchain-based platform will be utilized to track each 5G Edge node's (advertised) resources (e.g., spectrum resources and 5G Edge resources such SDN, NFV, and mobile edge servers) and update each nodes' reputation (e.g., persistent in offering resources). The BTCP uses the blockchain platform to log each node's actions/responses and update its reputation according to its participation. Most importantly, this decentralized trusted non-repudiating platform help in communicating computed results among the edge devices securely. This BTCP constitutes two primary components described in the following two sections.

## 4.1: A Lightweight Blockchain Platform:

The participant pool will be managed by a blockchain platform. While IoT devices open a new era for MEC services, particularly for a 5G Edge system, various trust issues are growing along with the benefits. The MEC using limited-resource (IoT) edge devices depends on the communication among the devices, as well as with the edge gateways and/or edge/cloud servers. The edge devices need to be authenticated to participate in the system, and they must trust each other. That is, the involved parties need to trust one another's statuses, e.g., identifiers, locations, and capabilities (e.g., available computing resources). A trusted third-party-based mechanism can provide a solution, but there is still the concern of trust, along with the issues like privacy and single point of failure.

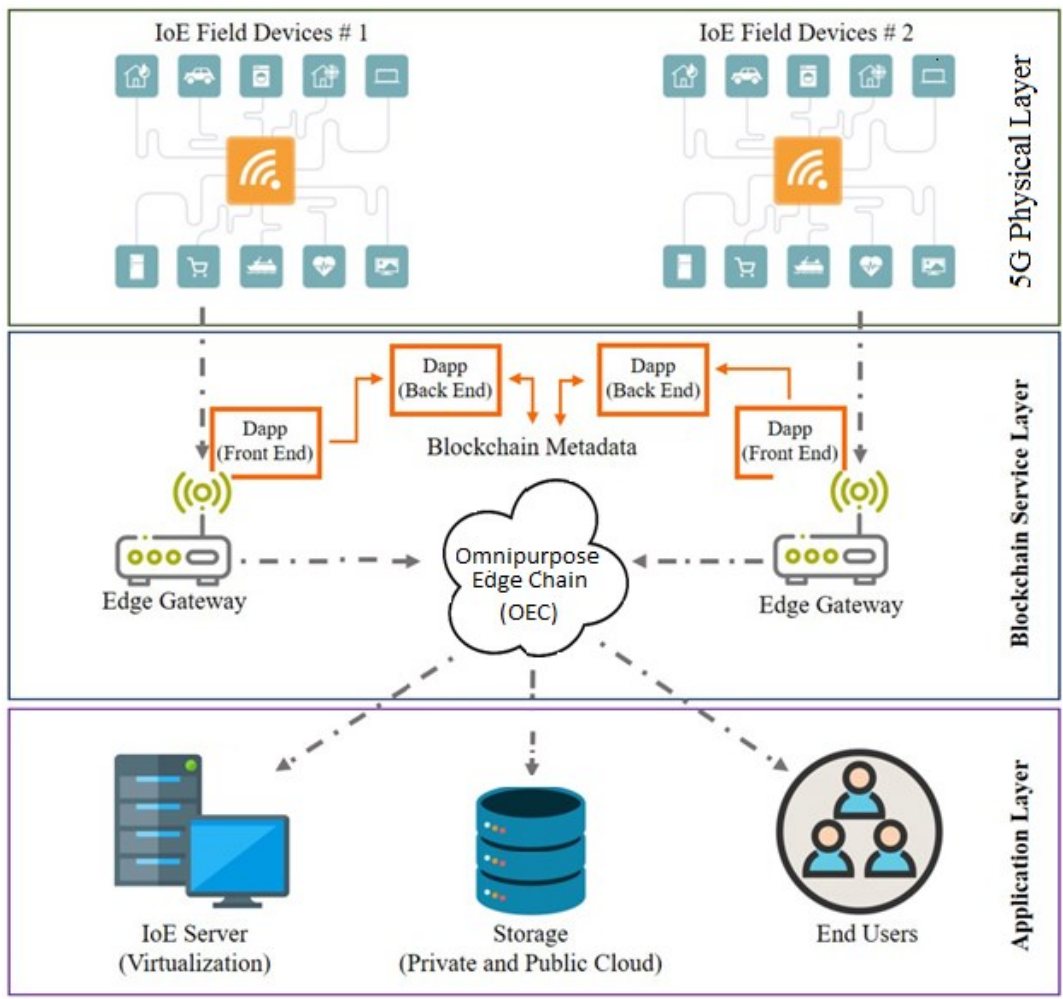

Fig. 1: The blockchain architecture through which edge devices are connected

There are existing works that have proposed solutions to some of these issues, such as authentication and privacy; however, the solutions have targeted mostly individual/specific problems that often cannot be applied for different settings, such as IoT device-based edge system computing. Therefore, a non-centralized and third-party-less framework is crucial for edge systems to provide secure, privacy-preserved trustworthy services. We propose to address this need by offering secure and trusted MEC services by utilizing the non-interactive ZKP technology to ensure authentication while participating nodes' privacy. We propose a blockchain architecture that leverages edge servers. We will call this system Omnipurpose Edge Chain (OEC). The edge servers and IoT gateways will be used to work as peer nodes of the OEC (*chaincode* implemented). We name them blockchain (BC) nodes. An edge/IoT device will be connected to a BC node.

Figure 1 shows a simplified diagram of how the edge devices will be connected to a Blockchain network. In this figure, OEC is the blockchain system, which has got several nodes. Edge devices get connected with any of the BC nodes, which are known as the corresponding node. The devices that are connected to the same corresponding node are called sibling nodes. Different BC nodes might have a different number of sibling nodes. The total number of nodes required might vary depending on the number of edge devices.

When an IoT device or an Edge server wants to subscribe to the MEC services, it will create its user profile. In its profile, the user will mention all necessary information, including its identity, device/resource information, and the payment detail (as applicable). Using a standard authentication mechanism, an IoT device will be connected to the Blockchain peer. Figure 2 shows a tentative protocol for how an IoT can subscribe to a peer of the OEC network. In the OEC network, there will be a peer who will work as an OEC manager. The OEC manager will be responsible for generating cryptographic keys. The OEC manager is randomly selected from the existing peers for a particular period.

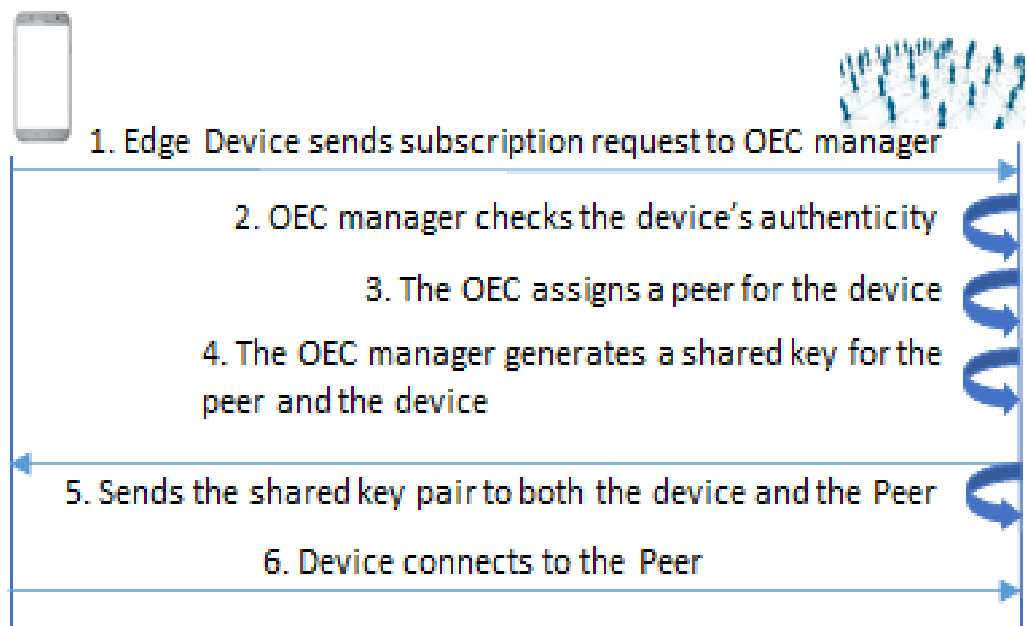

Fig. 2: A protocol for the subscription to OEC

In a permissionless blockchain setup, all transactions are broadcasted to each participant. The interactions between participating nodes are also visible to all participants. Hence, there exists no transaction privacy and confidentiality. Here, ZKP-based techniques can ensure these properties by letting others only know that a valid transaction has taken place, without sharing any information about the sender, recipient, and asset transferred. zk-SNARKS (zero-knowledge – Succinct Non-Interactive Argument of Knowledge) [54] is a popular ZKP technique. It is a verifier and prover-based technique, following the idea of defining a quadratic equation based on public data and private data (only known to prover) and generating proof for the verifier to validate. The process does not require any interaction/data-sharing between the prover and the verifier.

We consider the situation where the participating IoT/edge devices are not following a gateway-based architecture and thus are not connected to the powerful gateway, edge servers, and/or clouds, having a limited connection to the Internet so that they can access the blockchain as required. In the case of no Internet, these authentication/reputation models will allow participation in the pipeline, leaving verification and blockchain update processes to be performed once these nodes can connect to the Internet. It is worth mentioning that ZCash [55] is a cryptocurrency that implements zk-SNARKS to ensure the confidentiality of personal and transaction data. In a Z-to-Z transaction appearing on the public blockchain, it can be seen that the transaction took place and the fees were paid. However, the transaction amount, user addresses, and memo are encrypted, and hence not publicly visible. Currently, to make ZKP process short and efficient ZKP in a blockchain, a common reference string (public values) is generated and shared between the prover and verifier in the initial phase. Hence, if an adversary can access the secret randomness used for generating these parameters, he/she would be able to create false proofs for the verifier. To mitigate this problem, ZCash applies an elaborated multi-party ceremony to generate the public parameters, where multiple independent parties collaboratively create the parameters. This protocol ensures that to compromise the final parameters, all participants need to be dishonest or compromised. At a high level, zk-SNARKs works by converting what someone wants to prove into an equivalent form, demonstrating the knowledge of the solution. In this particular study, we aim to leverage the non-interactive ZKP designed for IoT [56].

**4.2. An Incentive-based Reputation Management:**

We propose to develop incentive-based reputation management on the blockchain platform to be implemented in the previous subtask. This management involves designing a protocol that is suitable for the dynamic formation of the edge system with heterogeneous nodes, as well as updating the system with various changes. These changes can be in the set of participants and the computing need. The protocol provides a payment system that incentivizes individuals according to their performance and commitment. This protocol also tracks the participants' behaviors and keeps the reputation updated accordingly.

The reputation management protocol considers the actions concerning forming or updating the dynamic MEC system. These actions primarily include "create," "join," "leave," and "break" the edge system. Note that to work with such a dynamic system, it is important to model the system with proper identification and property specification. For example, if a node is already subscribed/participating in an edge system, a condition can be applied to only allow a node to join another system after leaving the current one. A node cannot join a system if its advertised computing power is too low or its reputation score is too low. A node may not be able to join is if the edge system already has enough participants with sufficient computing and storage capacities to serve the system stakeholder's mission since an edge system does not need to be too large than the need. Most importantly, the stakeholder/leader (the creator node) of the edge system incentivizes the participating nodes with proper payments (maybe preliminarily advertised). This incentive is important to make the nodes motivated to share their resources by joining the edge system. The join action includes various lower-level actions to identify if everything is good to go. The payments for the joined/subscribed nodes according to their participation/performance in the MEC is made by the edge system's leader/creator node, and the reputation scores are updated accordingly. All these actions are recorded in the blockchain platform, including the payment/transactions. Frequently or as the stakeholder has configured/advertised, the incentives will be calculated and distributed. Now, whenever a node wishes to exit the system, it invokes the "leave" action. A user can leave a system to which it is currently subscribed. Once this "leave" process starts, each node should receive payment from the creator/leader node according to its unpaid service. Moreover, reputation scores will be updated (for the unjudged performance), and the blockchain will be updated with these scores/transactions. This task may invoke a series of low-level actions to safely exit.

To design this incentivization system, we require exploring different payment strategies and benefit models to find an incentive scheme. Any scheme developed must involve the participating nodes being treated fairly. Designing such a model considering nodes that participate partially, while the resources are heterogeneous, is not trivial. The overall goal of any member of the edge system is to earn (money) through their service. A simplified potential payment ($Z$) calculation for an edge device ($I$) can be done following Eq. 1:

$$Z_I = f(R, D_I, T_I) \times X_T \times \frac{f(R,D_I,T_I)}{g(R,P_I,S_I,C_I)} \qquad (1)$$

Here, $f(.)$ is a function that evaluates the computing performance (e.g., a scaled value) for task (computing requirement) $R$ based on the data ($D_I$) processed and time ($T_I$) required. This performance will calculate the monetary value of this participation. However, MEC with multiple nodes works based on coordination between the nodes, according to some schedule and dependency. In the cases when the node does not perform as advertised, this coordinated computing task will be hampered, which requires to be penalized. The last term of the above equation is simply doing it by taking the ratio of shown performance over the expected/advertised performance. The latter is calculated by function *g*(.) based on the task, the advertised processing ($P_l$), storage ($S_l$), and communication ($C_l$) capacities.

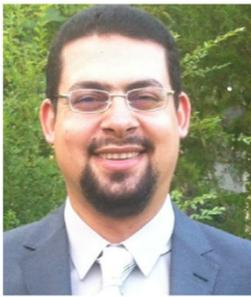
Hesham Abdelazim Ismail Mohamed (Hisham A. Kholidy) received his Ph.D. in Computer Science in a joint Ph.D. program between the University of Pisa in Italy and the University of Arizona in the USA. He works as an assistant professor at the Department of Networks and Computer Security (NCS), College of Engineering, State University of New York (SUNY) Polytechnic Institute. Before that, he worked as a postdoctoral associate at the University of Nevada, Reno, and Mississippi State University. During his Ph.D., he worked as an associate researcher at the NSF Cloud and Autonomic Computing Center, Electrical and Computer Engineering Dept. at the University of Arizona. He holds several patents in Cybersecurity published by the United States Patent and Trademark Office (USPTO), and he published more than 50 papers on high-quality journals and conferences. He was on the Technical Programs and Organization Committees for various IEEE and ACM conferences. He received several research awards such as the SUNY Poly dean's excellence research award, AFRL VFRA award, and SUNY Seed Research Award. He participated as PI, Co-PI, and senior personnel in several research projects and his research interests include Cybersecurity and SCADA systems security, 5G systems security, Cloud Computing and high-performance systems, service composition, and big data analytics.

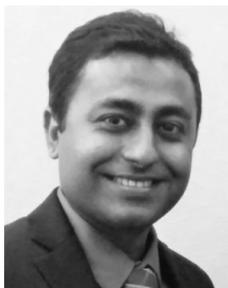
Mohammad Ashiqur Rahman (Member, IEEE) is an Assistant Professor with the Department of Electrical and Computer Engineering, Florida International University (FIU), USA. He is leading the Analytics for Cyber Defense (ACyD) Lab with FIU. Before joining FIU, he was an Assistant Professor with Tennessee Tech University. He received the Ph.D. degree in computing and information systems from the University of North Carolina at Charlotte in 2015. He has already authored or coauthored 75 peer-reviewed journals and conference papers. His research focus primarily includes artificial intelligence-based novel analytics design and development for network and information security, control-aware resiliency, and security hardening. He was on the Technical Programs and Organization Committees for various IEEE and ACM conferences.

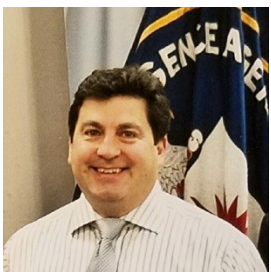
Experienced Technical Advisor with a demonstrated history of working in the Intelligence community and military industry. Skilled in U.S. Department of Defense, Intelligence community, Cyber Operations, Signals Intelligence, Communications Intelligence, Embedded Systems, Systems Engineering, Vulnerability Assessment, and Integration. Strong business development professional and senior program manager with a Master of Science (M.S.) focused in Electrical and Electronics Engineering from Syracuse University College of Engineering and Computer Science.

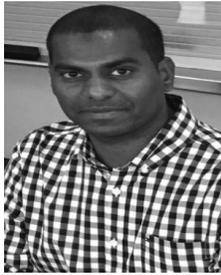
Zahid Akhtar (Senior Member, IEEE) received the Ph.D. degree in electronic and computer engineering from the University of Cagliari, Italy. He is currently an Assistant Professor with the Department of Network and Computer Security, State University of New York (SUNY) Polytechnic Institute, USA. Prior to that, he was a Research Assistant Professor with the University of Memphis, USA, and a Postdoctoral Fellow with the INRSEMT, University of Quebec, Canada, the University of Udine, Italy, Bahcesehir University, Turkey, and the University of Cagliari. His research interests include the computer vision and machine learning with applications to cybersecurity, biometrics, affect recognition, image and video processing, and audiovisual multimedia quality assessment.